\documentclass[12pt]{iopart}

\usepackage{graphicx}
\usepackage[square,sort&compress,numbers]{natbib}

\newcommand{\ff}{{\it ff}}
\renewcommand{\vec}[1]{\mathbf{#1}}
\newcommand{\uvec}[1]{{\hat{\textbf{#1}}}}
\newcommand{\fst}{FeSe$_x$Te$_{1-x}$}

\newcommand{\rhoff}{\rho_{\it ff}}

\renewcommand{\Re}{\textrm{Re}}

\newcommand{\eqref}[1]{(\ref{#1})}

\begin{document}

\title[Microwave Pinning in FeSeTe]{Pinning properties of FeSeTe thin film through multifrequency measurements of the surface impedance}

\author{N.~Pompeo$^1$, K.~Torokhtii$^1$, A.~Alimenti$^1$, G.~Sylva$^{2,3}$, V.~Braccini$^3$, E.~Silva$^1$}

\address{$^1$ Engineering Department, Università Roma Tre, Via Vito Volterra 62, 00136 Roma, Italy}
\address{$^2$ Physics Department, University of Genova, via Dodecaneso 33, 16146 Genova, Italy}
\address{$^3$ CNR-SPIN Genova, C.so F. M. Perrone, 24, 16152 Genova, Italy}
\ead{nicola.pompeo@uniroma3.it}
\vspace{10pt}
\begin{indented}
\item[]June 2020
\end{indented}

\begin{abstract}
We present high frequency measurements of the vortex dynamics of a \fst{} ($x=0.5$) thin film grown on a CaF$_2$ substrate and with a critical temperature $T_c\simeq18\;$K, performed by means of a dual frequency dielectric resonator at 16.4~GHz and 26.6~GHz. We extract and discuss various important vortex parameters related to the pinning properties of the sample, such as the characteristic frequency $\nu_c$, the pinning constant $k_p$ and the pinning barrier height $U$ relevant for creep phenomena. We find that the vortex system is in the single-vortex regime, and that pinning attains relatively high values in terms of $k_p$, indicating 
significant pinning at the high frequencies here studied. The pinning barrier energy $U$ is quite small and exhibits a non-monotonous temperature dependence with a maximum near 12~K. This result is discussed in terms of core pinning of  small portion of vortices of size $\propto\xi^3$ jumping out of the pinning wells over very small distances, a process which is favoured in the high frequency, short ranged vortex oscillations here explored. 
\end{abstract}


\section{Introduction}
The discovery of the new superconducting family of the Iron Based Superconductors (IBS) has sparkled a great research activity, both in light of their particularly intriguing peculiarities from a theoretical point of view \cite{Hosono2015} (competition between {antiferromagnetic} and superconducting order parameter, multiple bands), and from the point of view of applications, paramount among the others those concerning high currents and high fields, thanks to the high critical fields, lower anisotropy, robustness against disorder and lower vulnerability to misalignment in grain boundaries than their cuprate counterparts \cite{Hosono2017, Pallecchi2015}. 
Among IBS, \fst{} is particularly interesting, because of its simple crystal structure, the absence of the toxic Arsenic, and a very rich physics by itself. Indeed, in the mixed state vortex cores have shown discrete energy levels, pointing to Caroli-De Gennes-Matricon \cite{Chen2018} quantized energy levels and, as recently recognized, providing also a candidate for Majorana vortex modes \cite{HaoReview2018, Chiu2020}. 
Moreover, from a structural-compositional point of view, the difference between bulk and thin films properties \cite{Imai2016}, with the latter exhibiting enhanced superconducting properties down to the extreme situation given by single layer FeSe on SrTiO$_3$ \cite{Wang2012CPL}, further fuels research efforts. 
Focusing on the pinning properties, the importance of the substrate is also recognized in the studies aiming to improve the critical current density $J_c$, and to reduce the overall anisotropy of \fst{} films by investigating the effects of different substrates \cite{Tsukada2011, Braccini2013a}, and in the sensitivity of ion-implanted films on the position of the implantation layer with respect to the film-substrate interface \cite{Sylva_2018}. Relevant to the vortex matter is also the observed  ``fishtail'' peak effect \cite{Taen2009,Galluzzi2018}, intimately connected with the field and temperature evolution of the vortex system dynamics, which have been ascribed to several mechanisms in various superconducting materials (\cite{,Galluzzi2018} and references therein).
Within this panorama, microwave measurements have provided an important probe for thermal fluctuations in thin films \cite{Nabeshima2018}, in the determination of penetration depth and superfluid temperature dependence and the connected gap properties in single crystals \cite{Torsello2019,Okada2020} , in the study of various vortex parameters, such as the pinning frequency and the flux flow resistivity in single crystals \cite{Takahashi2012, Okada2012, Okada2014, Okada2015} {and thin films \cite{PompeoEUCAS2019}}.
Indeed, in the study of other - cuprate - superconductors, the dynamic regime explored at microwave frequencies allowed also to disentangle the ``intrinsic'', electronic mass anisotropy from the ``extrinsic'' one detected through $J_c$ measurements in d.c., affected by the action of pins of different dimensionalities \cite{Bartolome2019, Pompeo_2020}. Moreover, in combination - and in complementarity - with $J_c$ measurements, it allowed to highlight Mott-insulator-like effects in pinning, arising in d.c. regimes, in addition to the underlying pinning as detected through the high-frequency, short-range oscillations of vortices \cite{Pompeo2013}. 

Despite the versatility and richness of information that microwave studies allow to gather, to the best of our knowledge no such studies on the mixed state of \fst{} thin films have been published, apart from our previous preliminary report \cite{PompeoEUCAS2019}.
Hence, in this work, we present high frequency measurements of the mixed state of a \fst{} thin film performed by means of a dielectric-loaded resonator, capable of simultaneously operating at two distinct frequencies (16.4~GHz and 26.6~GHz). The well-established models for high-frequency vortex motion can be fully exploited with the aim of dual frequency measurements: we are able to reliably extract several vortex pinning properties, and to investigate their temperature (5~K-T$_c$) and field dependence (0-1.2)~T, with fields applied parallel to the sample $c$-axis. 

The structure of this paper is the following. In the following Section, the high frequency vortex motion models are presented, with a careful indication about their limits of applicability. In Section \ref{sec:experiment}, the sample deposition and properties, and the experimental technique, are presented. The experimental results are presented in Section \ref{sec:results} and discussed in Section \ref{sec:discussion}, respectively. Short conclusions are drawn in Section \ref{sec:conclusions}.

\section{High frequency vortex motion models}
\label{sec:models}
High frequency measurements  (i.e. in the microwave range, in this work) in the mixed state are advantageous since they probe several aspects of the vortex motion allowing to isolate each contribution \cite{PompeoLTP2020}. 
The microwave current density $\vec{J}$, oscillating with frequency $\nu$, exerts a time-varying Lorentz force $\vec{F}_L=\vec{J}\times\uvec{n}\Phi_0$ force per unit length on vortices ( $\Phi_0$ is the flux quantum and $\uvec{n}$ is the unit vector describing the orientation of the magnetic induction $\vec{B}=B\uvec{n}$). Vortices are thus set in a oscillatory motion around their equilibrium positions, from which they are displaced by $\vec{u}$. 
By moving, they experience a dissipative viscous drag $-\eta\vec{\dot{u}}$, microscopically related to the quasi-particles -- superfluid conversion \cite{Tinkham1996book}, and linked to the flux flow resistivity $\rho_{\ff}=\Phi_0B/\eta$. On the other hand, the short-range displacement from pinning centers determines the appearance of a pinning force well described as an elastic force $-k_p\vec{u}$ with pinning constant $k_p$ (also known as Labusch parameter), which measures the bottom curvature of the pinning wells \cite{Gittleman1966,Gittleman1968}. During the motion, stochastic processes (typically thermally activated but in principle also possibly originating from quantum tunneling processes) can allow vortices to detach from a pinning center - the so-called vortex ``creep''. The detection of these  processes allows to probe the height $U$ of the pinning barriers. 
By considering the force balance for a single vortex within a mean-field approach \cite{Golosovsky1996}, justified by the smallness of the oscillations with respect to the vortex spacing, one ultimately obtains the vortex motion complex resistivity $\rho_{vm}$  \cite{Coffey1991a, Brandt1992, Golosovsky1996, Pompeo2008}:
\begin{eqnarray}
\label{eq:rhovm}
\rho_{vm}&=\rhoff\frac{\chi+\rmi\nu/\nu_c}{1+\rmi\nu/\nu_c}
\end{eqnarray}
where $\chi\in[0,1]$ (with $\chi=0$ when no creep processes are present) is a dimensionless creep factor; $\nu_c$ is a characteristic frequency related to the well-known (de)pinning frequency $\nu_p=k_p/(2\pi\eta)$ \cite{Gittleman1966,Gittleman1968}, to which it coincides when $\chi=0$. These frequencies are particularly relevant in evaluating the high frequency performances of superconductors, marking the separation between the low-frequency $\nu\ll\nu_c$, Campbell regime, where pinning dominates in hindering vortex motion and dissipation is limited, and the high frequency $\nu\gg\nu_c$ regime, where the highly dissipative pure flux flow regime takes place. It is worth noting that it is possible to explore one or the other regime by changing the stimulus frequency, but still working with subcritical current densities, differently from d.c.
In the above, additional mechanisms like Hall effect, vortex mass (thought to be relevant at higher, near THz, frequencies \cite{Kopnin2002}) are neglected.
At this point it is important to stress that Equation \eqref{eq:rhovm}, with its specific prediction for the frequency dependence of $\rho_{vm}$, can be used quite in general and it is often validated by experiments, with particular significance of wideband  measurements \cite{Gittleman1966, Wu1995, Sarti2004, Silva2011, Silva2016}. On the other hand, the connection between the lumped parameters $\nu_c$ and $\chi$ and the specific microscopic properties of the underlying vortex system requires a further specialization of the model, with the consequent inevitable additional assumptions and limitations of applicability.
While for a short review of such models we refer to Ref. \cite{Pompeo2008}, we recall here the three most commonly used models, which will be used also in the analysis here proposed. The Gittleman-Rosenblum (GR) model \cite{Gittleman1966} neglects altogether creep phenomena, setting $\chi=0$ and thus $\nu_c=\nu_p$. The Coffey-Clem \cite{Coffey1991a, Clem1992a, Coffey1992} and Brandt \cite{Brandt1992} models, on the other hand, take into account creep, following different approaches.
The former assumes periodic, sinusoidal pinning potential that enables it to describe thermal creep in the whole range of creep factors between 0 and 1, the upper limit corresponding to a total detachment of vortices from pins yielding pure flux flow motion. The latter incorporates creep by assuming a time-relaxing pinning constant: although the dynamics of possible $\chi$ values is limited to $\chi \leq0.5$, this approach allows in principle to generalize creep including quantum processes.

Having discussed $\rho_{vm}$, a last step is needed to connect it to the complex resistivity $\tilde\rho$ measured in the experiments. The microwave currents which excite vortex motion are sustained by both quasi-particles and superfluid: within the two fluid model, their conductivities, $\sigma_s$ and $\sigma_n$ sum up as $\sigma_s+\sigma_n=\sigma_1-\rmi\sigma_2$ and, once coupled back with the moving vortices, yield an overall resistivity $\tilde\rho$ as follows \cite{Coffey1991a}:
\begin{eqnarray}
\label{eq:rhotilde}
\tilde\rho=\frac{\rho_{vm}+\rmi/\sigma_2}{1+\rmi\sigma_1/\sigma_2}
\end{eqnarray}
Since the two-fluid conductivity is a small contribution to the field-dependent response, at least at fields and temperatures not too close to the transition, it is often possible to relate the experimental $\tilde\rho$ to $\rho_{vm}$ only, as better described in Sec.\ref{sec:experiment}.

In the next Section we present some sample details and the microwave experimental technique used for the measurement of $\tilde\rho$ and $\rho_{vm}$.

\section{The experiment}
\label{sec:experiment}

\fst{} thin films of 240~nm thickness were deposited on $7\times7\;$mm$^2$ CaF$_2$ single crystals in a high vacuum PLD system equipped with a Nd:YAG laser at 1024~nm, using a FeSe$_{0.5}$Te$_{0.5}$ target synthesized with a two-step method \cite{Palenzona2012}. The optimized laser parameters to obtain high quality epitaxial 11 thin films \cite{Braccini2013} were a 3~Hz repetition rate, a 2~J/cm$^2$ laser fluency (2~mm$^2$ spot size) and a 5~cm distance between target and sample. The deposition was carried out at a residual gas pressure of 10$^{-8}\;$mbar while the substrate was kept at a temperature of 300~$^\circ$C.

Surface impedance $Z$ \cite{Collin1992} in the microwave range has been measured by means of a cylindrical dielectric-loaded electromagnetic resonator \cite{Pompeo2014,Alimenti2019a}, with a coaxial, cylindrical single crystal sapphire rod. 
A thin metal mask 
was used to preserve the cylindrical symmetry of the region exposed to the electromagnetic fields.

The resonator operates simultaneously at two modes, TE$_{011}$ and TE$_{021}$, with resonant frequencies $\nu_1=16.4\;$GHz and $\nu_2=26.6\;$GHz, respectively, selected since they induce microwave currents parallel to the sample surface and because of the relatively high sensitivity which they confer. 
A Vector Network Analyser provides the measurements of the two-port scattering coefficients of the resonator. Frequency sweeps yield the quality factors $Q_i$ and $\nu_i$ ($i=1,2$) of both modes by means of complex fits of the nonideal scattering coefficients \cite{Pompeo2017a,Torokhtii2019a}.

The resonator is placed in He-flow cryostat, which allows to control the temperature down to $5\;$K with a $\pm0.01\;$K stability. The cryostat is placed in the bore of a conventional electromagnet, capable of delivering static magnetic fields $\mu_0H$ perpendicularly to the sample surface up to 1.2 T. 
The surface impedance $Z=R+\rmi X$ of the sample, where $R$ and $X$ are the surface resistance and reactance, respectively \cite{Collin1992}, is extracted from $Q$ and $\nu$ through the standard relationship \cite{Staelin1994book}:
\begin{eqnarray}
\label{eq:QZ}
R(T,H)+\rmi\Delta X(T,H)=G\frac{1}{Q(T,H)}-\rmi2G\left(\frac{\nu(T,H)}{\nu_{ref}}-1\right)-{\rm bck}
\end{eqnarray}
where $\Delta$ denotes a variation with respect to a reference values (typically for a given $T$ or $H$, to which 
$\nu_{ref}$ corresponds); $G$ is a mode-dependent numerically computed geometrical factor and ``bck'' is a field-independent background. 
In the local limit, the complex resistivity $\tilde\rho$ can be obtained as $Z=\sqrt{\rmi\mu_02\pi\nu\tilde\rho}$ in thick samples \cite{Collin1992}, i.e. having thickness $d\gg{\rm max}(\delta, \lambda)$ where $\delta$ and $\lambda$ are the normal skin depth and London penetration depth, respectively, while $Z=\tilde\rho/d$ in the so-called thin film approximation, when $d\ll{\rm max}(\delta, \lambda)$ \cite{PompeoImeko2017a}.  

By applying a magnetic field $H$, since not too close to $H_{c2}$ and $T_c$, pair-breaking effects are negligible and $\sigma_1/\sigma_2\ll1$, following Eq.\eqref{eq:rhotilde} one derives
$\Delta Z(H,T)=Z(H,T)-Z(0,T)=\Delta\tilde\rho(H)/d\simeq\rho_{vm}(B)/d$ (where in the last {approximate} equality the London limit is considered $B\simeq \mu_0 H$): thus, by performing field sweeps at fixed temperature $T$, the vortex motion resistivity can be directly determined.

A last step in the determination of the vortex parameters is the following. Thanks to the dual frequency operation of the measurement cell, for each measurement point at a given $T$ and $H$ intensity, four observables are available, i.e. the real and imaginary parts of the vortex motion resistivity $\rho_{vm}=\Delta Z\cdot d$ at the two measuring frequencies $\nu_1$ and $\nu_2$. The three vortex parameters $\rhoff$, $\chi$ and $\nu_c$ appearing in Eq. \eqref{eq:rhovm} can thus be extracted by a closed-from analytical inversion. Actually a forth unknown can be also determined: as a check of the entire procedure, the ratio $G_2/G_1$ (see Eq.\eqref{eq:QZ}) of the geometrical factors is determined from the inversion of the equations, and then compared with the nominal value.

\section{Experimental Results}
\label{sec:results}

In the following we present measurements performed on the \fst{} film, with particular focus on the filed-dependent properties. The quasi-static magnetic field $H$ is applied normally to the sample surface, i.e. parallel to its $c$-axis, at fixed $T$ (field sweeps).

\begin{figure}[htbp]
\centering
\includegraphics[height=6cm]{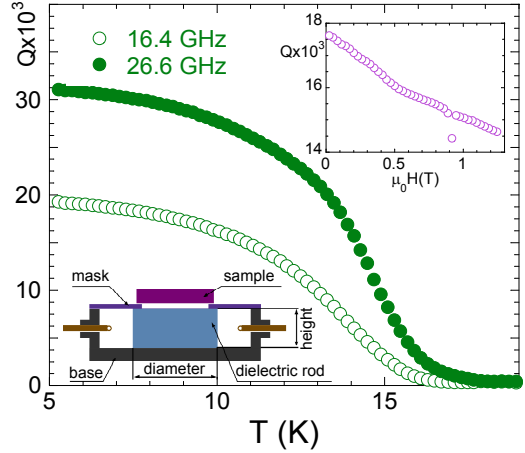} 
\caption{$Q_i(T)$ at zero field for the two resonant modes. Upper inset: $Q_1(H)$ at fixed $T=12\;$K, showing an anomaly around 0.9~T due to magnet power supply issue: these anomalies extend for a few mT and are removed in the subsequent presentation. Lower inset: sketch of the dielectric resonator.}
\label{fig:Q(T)}
\end{figure}

Figure \ref{fig:Q(T)} reports the resistive transition, measured in zero field cooling (ZFC) conditions, in terms of the resonator unloaded quality factor $Q$. 
The quality factors $Q_i(T)$ do not show any signature of second phases, and no evident signs of weak-link behaviour or of serious inhomogeneities, that would appear as a ``bump'' in the transition or in a very broad, nonsaturating temperature dependence. $T_c\sim18\;$K can be estimated from mode 2, which has a higher sensitivity in the highly dissipative state close to $T_c$: this value is in good agreement with those obtained through d.c. measurements on similar samples \cite{Braccini2013}. Exploiting the real part of Eq.\eqref{eq:QZ} for $Q_2$ data above $T_c$, and by properly estimating the resonator background term $\Re({\rm bck})$ in  Eq.\eqref{eq:QZ} \cite{Alimenti2019a}, the normal state resistance $R_n$ and thus the normal state resistivity $\rho_n=R_n\cdot d\sim(3.0\pm0.2)\cdot10^{-6}\;{\rm \Omega m}$ can be estimated, in good agreement with d.c. measurements on similar films, and literature data on other \fst{} thin films \cite{PompeoEUCAS2019}, polycrystals \cite{Li2015} and  single crystals of similar composition \cite{Okada2015}. Moreover, $\rho_n$ allows to compute the skin depth $\delta=\sqrt{\rho_n/(\pi\nu\mu_0)}\sim6.8\;{\mu\rm m}$ and $5.4\;{\mu\rm m}$ (for mode 1 and 2, respectively), both yielding $\delta\gg d$, thus self-consistently ensuring the applicability of the thin film approximation. 

\begin{figure}[htbp]
\centering
\includegraphics[width=0.95\columnwidth,height=12cm]{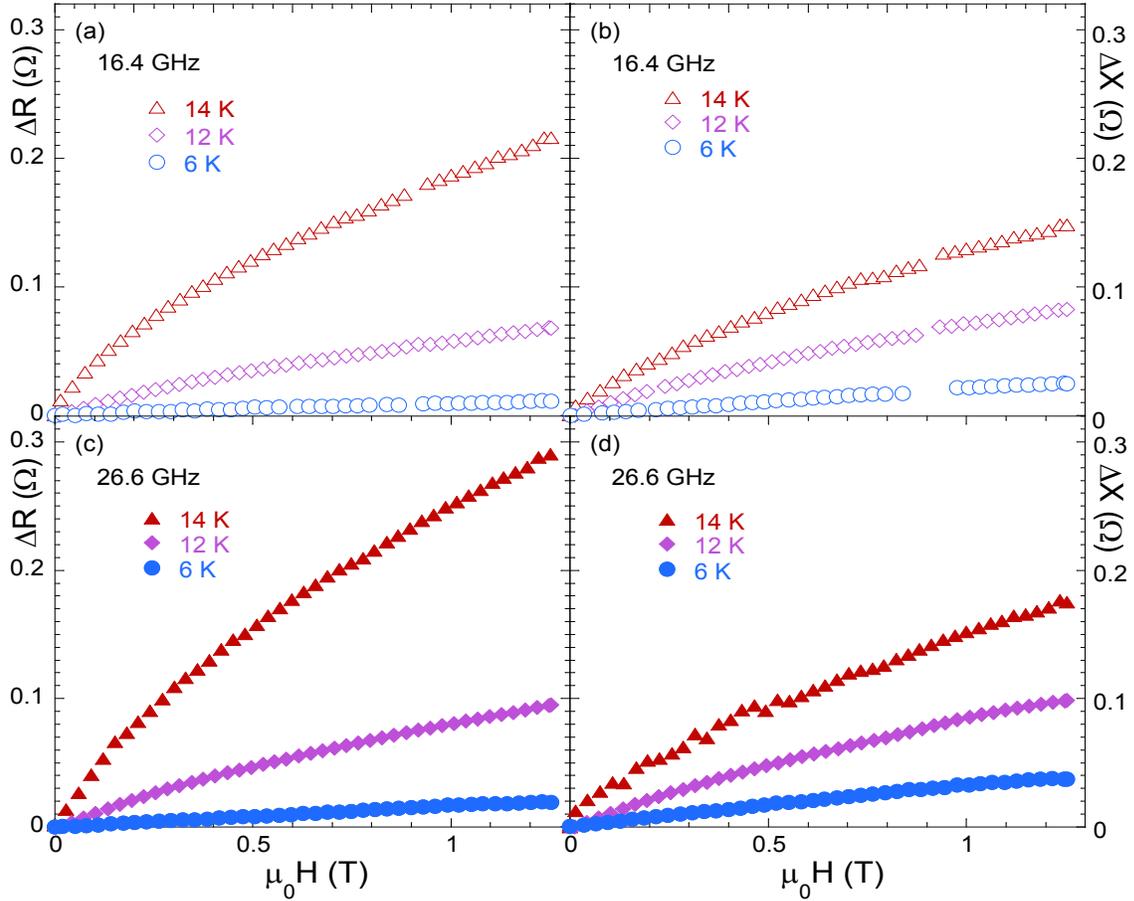}
\caption{{Surface resistance $\Delta R$ - panels (a) and (c) - and reactance $\Delta X$ - panels (b) and (d) - vs field $H$ for the two frequencies, at selected $T$.}}
\label{fig:Z(H)}
\end{figure}

We now consider field sweep measurements, obtained at fixed temperatures by sweeping the field from zero to the maximum value $\sim1.2\;$T after a zero-field cooling to the desired temperature. 
In Fig. \ref{fig:Z(H)} the surface resistance $\Delta R(H)=R(H)-R(0)$ and reactance $\Delta X(H)=X(H)-X(0)$ field-induced variations are shown for both modes at selected temperatures.
Several general features can be observed: both $\Delta R$ and $\Delta X$ increase with the applied field intensity, exhibiting a slight downward curvature. By increasing $T$, $\Delta R$, and thus dissipation, increases as expected by approaching $T_c$. Moreover, no steep rise is detected in the low field region, which would be a signature of weak links with the related Josephson or Abrikosov-Josephson vortices \cite{Silva1990, Gurevich1993, Pompeo2019a}, pointing to a good connectivity of the film. 

For further qualitative and quantitative discussion, the vortex parameters need to be extracted. Hence, in the next Section we analyse the data in the framework of the vortex motion models.

\section{Discussion}
\label{sec:discussion}

By resorting to Eq. \eqref{eq:rhovm} and combining the measurements taken at the two frequencies for each field sweep, the vortex parameters $\rhoff$, $\nu_c$ and $\chi$ are extracted. The previously mentioned check on the so-extracted $G_2/G_1$ ratio has also been performed, yielding $G_2/G_1=5.6\pm0.5$ (standard deviation is reported), in very good agreement with the computed value $G_2/G_1=5.58$. The relatively small dispersion can be considered, at the same time, as a confirmation of the applicability of the model to the data (i.e., the model being capable of capturing the main features of the measurements) and as an estimation of uncertainty on the parameters thus extracted.

\begin{figure}[htbp]
\centering
\includegraphics[width=0.5\columnwidth]{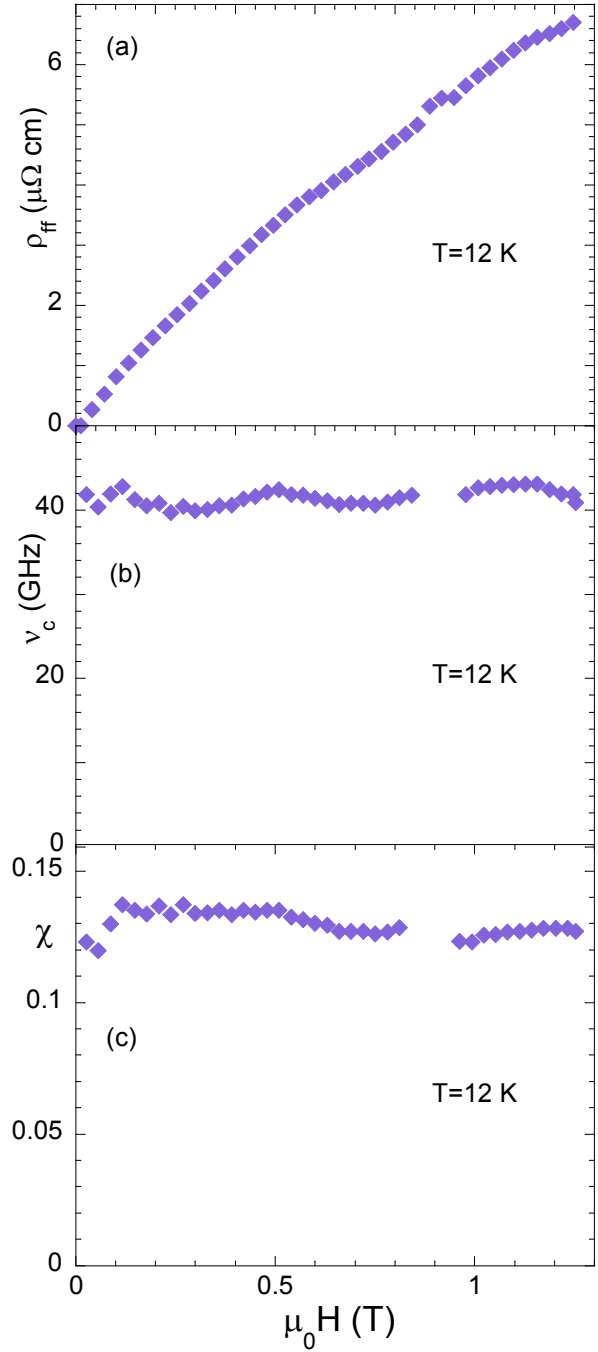}
\caption{Vortex motion parameters extracted through analytical inversion at $12\;$K.}
\label{fig:vmp(12K)}
\end{figure}

A representative example of the obtained vortex parameters is reported in Fig. \ref{fig:vmp(12K)} for $T=12\;$K.
Several observations can be made. The flux flow exhibits a slight downward curvature. 
Quantitatively, the slope of $\rhoff(B)$ for $B\rightarrow0$  can be extracted and compared to the expected dependence $\rhoff=\alpha\rho_n B/B_{c2}$, where the adimensional slope $\alpha$ is connected to the microscopic properties of the superconductor, in terms of the nature of the quasi-particles density of states \cite{Caroli1964,Blatter1994}
and of the features of the superconducting gap, including the extension to anisotropic \cite{Kopnin1997} and multiple gaps \cite{Shibata2003, Okada2015, Silaev2016}. 
By roughly estimating $B_{c2}\sim37\;$T at 12~K from literature data \cite{Bellingeri2014}, $\alpha(B\rightarrow0)\sim1$, smaller than the values $>1$ expected in multiple gap superconductors and actually observed also in other \fst{} thin film \cite{PompeoEUCAS2019}, and larger than what detected in single crystals \cite{Okada2015}. These puzzling results, 
the study of $\alpha$, the whole $B$-dependence {of $\rho_{ff}$} and its evolution with temperature is outside the scope of this work, and is postponed to future in-depth analyses. From now on, we focus exclusively on the pinning quantities.

\begin{figure}[htbp]
\centering
\includegraphics[width=0.5\columnwidth]{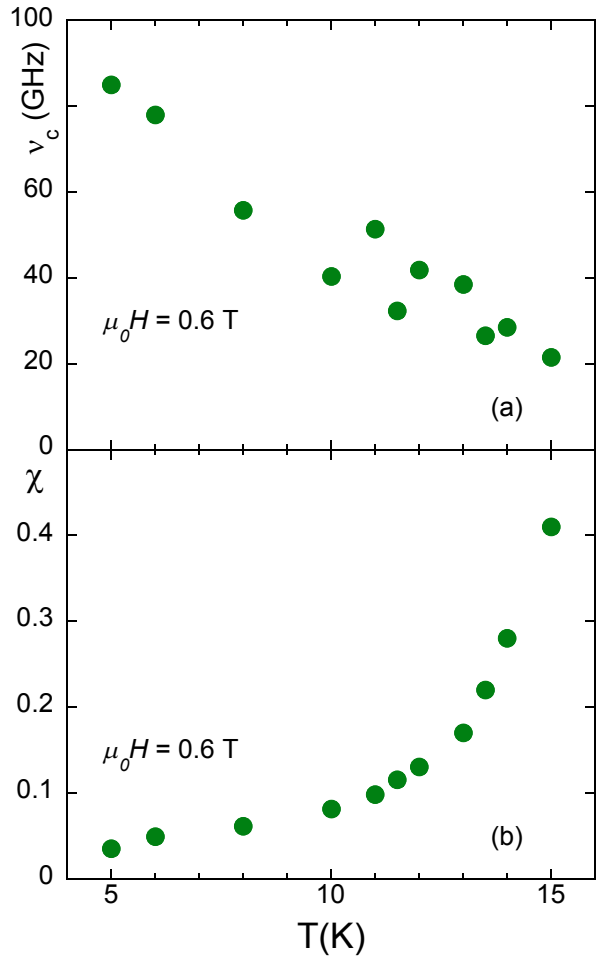}
\caption{Temperature dependence of the characteristic frequency $\nu_c$ - panel (a) -  and of the creep factor $\chi$ - panel (b) - at 0.6~T.}
\label{fig:pin(T)}
\end{figure}

The pinning related quantities $\nu_c$ and $\chi$ are remarkably flat with respect to the field, for all the $T$ investigated. This field-independence, which will reflect itself on the quantities which will be derived in the following analysis step, is a clear indication that the vortex system is, in the field range here considered, in a single-vortex regime, as it will be discussed further on. This regime, at least for relatively low fields, is often realized at high frequencies, even if in the same $T$-$H$ region d.c. measurements exhibit collective behaviour of the vortex system. This is due
to the very different dynamics: very small vortex oscillations take very short time intervals, so that vortices move essentially in place without significantly varying their interaction with neighbouring vortices during their motion.
Since the field dependence of the vortex parameters $\nu_c$ and $\chi$ is nearly constant, {in Fig. \ref{fig:pin(T)} we plot the data taken at $\mu_0H=0.6\;$T, at all the temperatures investigated,} as a representative dependence of the same parameters on the temperature.
It can be seen that $\nu_c$ has, as it could be expected, an increasing trend by lowering the temperature, pointing to a more efficient pinning action on cooling. As far as the absolute values are concerned, to the best of our knowledge no determinations of this characteristic frequency are available in the literature,  because of the scarcity of microwave studies on IBS in general and \fst{} systems in particular, and because
{more common single frequency measurements only allow to estimate $\nu_p$, and not $\nu_c$, with the strong assumption of zero/negligible creep \cite{Gittleman1966}. 
A value $\nu_p=20\;$GHz at $\mu_0H=0.5\;$T and $t=T/T_c=0.66$ has been found on a \fst{} thin film \cite{PompeoEUCAS2019} with similar $T_c$; $\nu_p=16\;$GHz at 1~T and $t=0.71$ ($T_c\sim14\;$K) on a FeSe$_{0.4}$Te$_{0.6}$ single crystal \cite{Okada2015}, and lower values in other iron superconductors \cite{Narduzzo2008, Okada2012, Okada2013} were also estimated. 
The pinning frequency $\nu_p$ as derived from the zero-creep model is actually an underestimation of the real quantity, as extensively discussed in \cite{Pompeo2008, PompeoIMEKO2020}, but the obtained very large values of $\nu_c$ are a clear indication of possible good pinning performance of the sample under study. 

The creep factor decreases with decreasing temperature, analogously indicating a decrease of the thermal depinning from pins. Whether this behaviour results by a reduction of the available thermal energy or an increase of the pinning efficiency, it cannot be discerned at this stage.

At this point of the discussion we have extracted as much information as possible by remaining in the framework of a general model, represented by Eq. \eqref{eq:rhovm} for $\rho_{vm}$. Indeed, it does not make specific assumptions about the nature of creep or the shape of the pinning potential, providing only a specific prescription for the frequency dependence which, as the above results have shown, well reproduces the experimental data.
In order to discuss in more depth the physics of pinning at very-short-range vortex displacement, specific models have to be used. 

Both the CC and the Brandt models, recalled in Section \ref{sec:models}, allow to extract a barrier energy value $U$ from the creep factor $\chi$ and the pinning constant $k_p$ from $\nu_c$ (in combination with $\rhoff$ and $\chi$), unfortunately yielding numerically different results for the vortex parameters. We anticipate that (i) our results do not allow to discriminate which model is most suitable, so that we conservatively discuss both of them but (ii) interestingly enough, many physical considerations hold and are unaffected by the model chosen, thus enforcing the generality of the results themselves.

We discuss first the pinning constant (Labusch parameter) $k_p$ as calculated from the data by inversion. The obtained curves for $k_p$  vs the field are almost field-independent  as reported in Fig. \ref{fig:kpH} for the CC model as an example), with only a slight increase with the field. 
\begin{figure}[htbp]
\centering
\includegraphics[width=0.5\columnwidth]{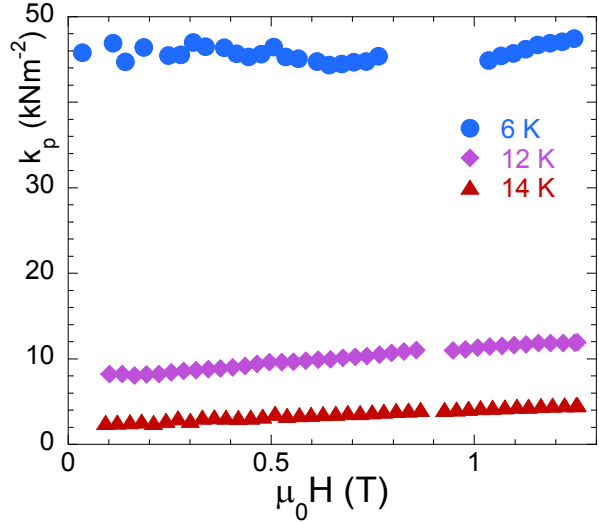}
\caption{Pinning constant $k_p$ vs $H$ at selected temperatures.}
\label{fig:kpH}
\end{figure}
The values at $\mu_0H=0.6\;$T as a function of the temperature are reported in Fig. \ref{fig:kpt}, for both the CC and Brandt models, together with the values obtained through the GR model calculated for mode~1 for comparison, which yields the smallest values possible compatible with the data if only one measurement frequency were considered \cite{Pompeo2008}. 
The pinning constant $k_p$ from the Brandt model is close to the GR values, as expected and extensively discussed in \cite{Pompeo2008}. 
It is important to note that the pinning constant presents only a small dispersion among the various models (which can be taken as an estimation of the uncertainty on the measured value $k_p$), thus exhibiting a substantial model independence.
This gives confidence on the robustness of the analysis proposed in the following. 
\begin{figure}[htbp]
\centering
\includegraphics[width=0.5\columnwidth]{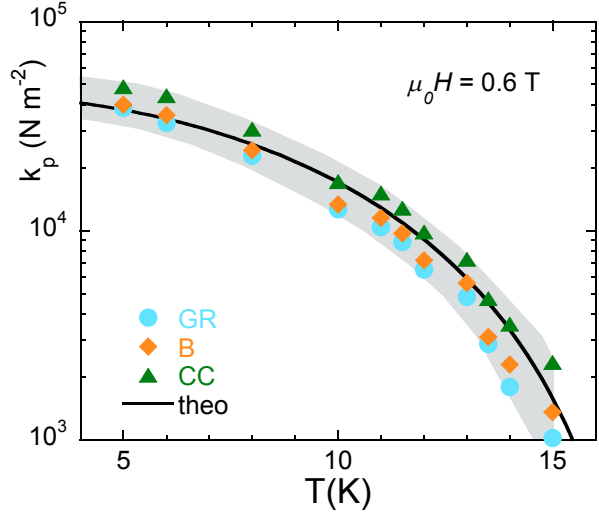}
\caption{{Temperature dependence of the pinning constant $k_p$ for the CC, B and GR models together with theoretical $T$ dependence (see text) at 0.6~T.}}
\label{fig:kpt} 
\end{figure}
It can be noted the the $T$-dependence of $k_p$ does not show any hint to saturation by lowering $T$. 
This feature has a very intriguing connection with the microscopic state of \fst{}. In fact, ultimately $k_p$ is proportional to the condensation energy, since it depends on the energy gain of a vortex occupying a defect. We start by taking a simplifying approximation where the elastic energy per unit length $k_p\xi^2/2$ of a detached vortex is  equal to the condensation energy per unit length $\mu_0 H_c^2 \xi^2\pi/2$ for a defect of width $2\xi$ \cite{Golosovsky1996, Pompeo2014a}, taking for the thermodynamic field $H_c=\sqrt{H_{c1}H_{c2}}$, with $H_{c1}=\Phi_0/(4\pi\Phi_0\lambda^2)$. Using $H_{c2}(T)$ from Ref. \cite{Tarantini2011},  determined on \fst{} thin films grown on LaAlO$_3$, and $\lambda(T)$ from single crystal measurements \cite{Okada2015}, the obtained temperature dependence of $k_p$ is consistently recovered. Considering also the absolute values for {$\mu_0 H_{c2}(0)\sim53\;$T} and $\lambda(0)\simeq520\;$nm from the above references, only an additional scale factor $\sim0.6$ is necessary to recover our result, indicating a  quite strong pinning which, within a crude picture, would be equivalent to a 60\% of the vortex line effectively pinned. This point will be further discussed in the following. Indeed, focusing on the absolute values of $k_p$, they reach $50\;$kN/m$^2$ at the lowest $T$. For comparison, the values attained near 10 K on YBa$_2$Cu$_3$O$_{7-\delta}$, one of the cuprates more mature technologically, are as high as $300\;$kN/m$^2$  at sub-THz frequencies \cite{Parks1995} and $150\;$kN/m$^2$ \cite{YBCOhighFields2017} in the same frequency range as the measurements here presented. 
Considering that \fst{} is far from being optimized, future improvements could further reduce the difference. 
Further considerations can be done through a comparison with the d.c. critical current density $J_c$. In order to obtain comparable quantities, one can equate the maximum pinning force $k_p \xi$ exerted by a pin at the maximum distance given by the coherence length $\xi$ \cite{Koshelev1991, Golosovsky1996}, to the Lorentz force $J_c\Phi_0$ exerted by the critical current density. In this way, a ``microwave'' estimation of the critical current $J_{c,mw}\sim k_p\xi/\Phi_0$ is obtained. 
By crudely computing $\xi(t=0.66)=\xi_0/\sqrt{1-t^2}=2.0\;$nm from $\xi_0=1.5\;$nm \cite{Pallecchi2012}, one gets $J_{c,mw}\sim800\;$kA/cm$^2$, to be compared with the d.c. value $J_c\sim$ 300-400~kA/cm$^2$ measured on similar samples in a similar temperature-field range \cite{Braccini2013a}.
Taking into account the possible variability in the pinning potential from sample to sample, our result for pinning do not appear far from the results obtained by the d.c. technique.

We note that the values for $k_p$ of the present samples are almost an order of magnitude larger than those extracted through a single frequency GR analysis, reported in \cite{PompeoEUCAS2019}. Although the two-frequency approach is more reliable, the order of magnitude difference can only indicate that the shape of the pinning potential is still extremely sample-dependent. Whether it is simply due to a different kind and concentration of pinning points or to more basic properties related to the exotic, multigap-related properties of the vortex matter, it remains to be assessed. However, for what concerns $k_p$, the present results indicate that the \fst{} sample grown on a CaF$_2$, substrate known to offer interesting improvements in the critical current density $J_c$, is also capable of delivering significant pinning in the high frequency regime with very-short-range vortex oscillations. 

We now consider the pinning barrier energy $U$ extracted again within both the CC and Brandt models from the creep $\chi$ parameter. 
Fig. \ref{fig:ut} reports $U(T)$ at $\mu_0H=$0.6~T, without loss of generality due to the nearly field-independence of $U$ (since it is derived from $\chi$) at fixed $T$.
\begin{figure}[htbp]
\centering
\includegraphics[width=0.5\columnwidth]{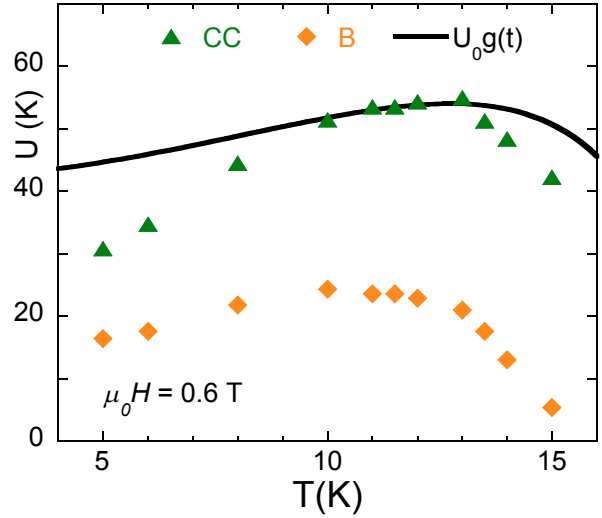}
\caption{{Temperature dependence of the pinning barrier energy $U$ for the CC and B together with theoretical $T$ dependence (see text) at 0.6~T.}}
\label{fig:ut} 
\end{figure}
As for $k_p$, the temperature trend is model-independent, so that the same considerations hold, while the absolute values differ for roughly a scale factor of 2, which is intimately linked to the different assumptions of the models \cite{Pompeo2008} (creep above sinusoidal pinning potential in the CC mode, thermal relaxation of the pinning constant in the Brandt model).
Differently from $\chi$, $U$ exhibits a non-monotonous $T$ evolution, with a hump around 12~K. This feature has been observed in several experiments of magnetic relaxation \cite{Leo_2015, Galluzzi_2019}, also in other materials \cite{Tahan_2011}, and of magnetoresistivity \cite{Leo_2015, Zhaofeng2016}. 

An analytical expression $g(t)$ for the $T$ dependence of $U$ is recalled in \cite{Budhani1990}, a generalization of the approach proposed by Tinkham in \cite{Tinkham1988}, and used in \cite{Leo_2015} to analyse the peak observed in a FeSe sample. Whereas in \cite{Tinkham1988}, $g(t)$ is derived assuming that the jumping volume has transversal (with respect to the field orientation) size of the length scale given by the vortices spacing $a=\sqrt{\Phi_0/B}$ and longitudinal length scale $\xi$, in \cite{Budhani1990} a generalization to jumping correlated volumes of different dimensionalities $\xi^n$ is considered. 
In particular, $\xi^3$ implies small portions jump across dense pinning points within the same lattice cell.
Starting from a $T$ dependence for the thermodynamic field $H_c\propto(1-t^2)$ and $\xi\propto((1-t^2)/(1+t^2))^{1/2}$, $g(t)=(1-t^2)^2[(1+t^2)/(1-t^2)]^{n/2}$. A non-monotonous $T$ dependence as the one experimentally observed is obtained only for $n=3$ (plotted in Fig. \ref{fig:ut} as a continuous line, and scaled through a properly chosen $U_0$), meaning that the jumping vortex portions are extended in all the three spatial dimensions along the same length scale of $\xi$, i.e. they are essentially 3D and thus small portions. 
Taking into account the small vortex displacement under the microwave current, a possible scenario for the pinning potential envisions closely packed pinning sites, separated by an average distance $\xi<l<a$ {(with a maximum $\xi(15\;{\rm K})=2.7\;$nm and minimum $a(1.2\;{\rm T})=45\;$nm)}, responsible for short-ranged jumps of small ($\xi^3$) vortex pieces. Due to the short distance between pinning sites, the corresponding energy barriers should be also small. 
This is consistent also with the small values obtained for $U$, lower than the values observed with  other techniques \cite{Galluzzi_2019, Bellingeri2012a, Bellingeri2014, Leo_2014, Leo_2015, Zhaofeng2016}, by a factor 5-10. This fact is intrinsically linked to the difference in the techniques: at microwaves, vortices are displaced from their equilibrium positions by very small amounts in very short times. Thus, they do not gain enough energy to overcome large barriers, but are able to ``jump'' only over the smallest barrier heights. In other words, thinking to a pinning potential profile with many barriers different in widths and heights \cite{Kierfeld2004}, microwaves are sensitive only to the smallest. 
Hence, this fact reinforces the above scenario of small $\xi^3$ size of the jumping portion of the vortices
\footnote{Another common interpretation \cite{Tahan_2011, Galluzzi_2019} of the maximum in $U(T)$ is a crossover from low-$T$ elastic creep to a high-$T$ plastic creep. Given the single vortex regime in the whole $H-T$ range here explored, this does not seem a feasible explanation for our results.}. 
It would be interesting to determine if in samples with extended defects (as those obtained with ion irradiation \cite{Masseee1500033}), a creep connected to extended pins would appear ($n<3$) or if also in that pinning landscape microwaves would sense only jumping volumes $\propto\xi^3$.
As a last observation, 
$U$ decreases with lowering $T$ faster than the theoretical predictions, but consistently with other results reported in literature \cite{Leo_2015}. The extrapolation to zero temperature is not sufficiently reliable to recognize possible signatures of quantum creep \cite{Klein2014}, which would appear as artifact $U(T\rightarrow0)\rightarrow0$, corresponding to finite creep even at zero temperature.

\section{Conclusions}
\label{sec:conclusions}

We have measured the microwave surface impedance of a \fst{} thin film grown on CaF$_2$ with $T_c\sim18\;$K.
No signatures of intergrain weak links have been observed, pointing to an overall good inter-connectivity of the film.
By applying a static magnetic field parallel to the film $c$-axis, the vortex motion complex resistivity have been measured through a dielectric resonator operating simultaneously at two frequencies. 
By exploiting the dual frequency measurements, the frequency dependence of the vortex motion complex resistivity has been exploited and the relevant vortex parameters extracted. It comes out that in the field range explored, the pinning-related quantities are field independent, an indication of a single vortex regime. The pinning constant, a measure of the curvature of the pinning wells near the bottom, 
exhibits values lower than those of the technologically mature YBa$_2$Cu$_3$O$_{7-\delta}$, so that future improvements could reduce the difference. 
Its temperature dependence shows no signs to saturate at low (down to 5 K) temperatures, similarly to the superfluid temperature dependence of these compounds, which is considered as a signature of their multigap nature.
The creep factor, related to the pinning barrier height, has also been determined.
The resulting pinning barrier height exhibits a non-monotonous trend, with a wide maximum at 12~K, associated with 3D small vortex volume involved in the thermal jumps. This is in agreement with observations obtained by means of different experimental techniques in much different dynamic regimes, i.e. d.c. magnetoresistivity or magnetic relaxation. 
Future works could address the angular dependence of the vortex parameters, in order to confirm the 3D nature of the vortex pinning and creep.

\section*{Acknowledgments}
Work supported by MIUR-PRIN project “HiBiSCUS" - grant no. 201785KWLE.
The authors acknowledge A. Provino and P. Manfrinetti for the target preparation, and M. Putti for scientific discussion.

\providecommand{\newblock}{}

\end{document}